\documentclass[aps,prd, superscriptaddress, preprintnumbers,showpacs]{revtex4}
\usepackage{epsfig}
\usepackage{graphicx}
\usepackage{amsmath}
\usepackage{bm}
\usepackage{amssymb}
\usepackage{color}

\newif\ifold             \oldtrue            
\def\ba{\begin{eqnarray}}
\def\ea{\end{eqnarray}}

\newcommand{\be}{\begin{equation}}
\newcommand{\ee}{\end{equation}}

\begin{document}

\title{Magnetic field driven instability  in planar NJL model in real-time formalism}

\author{O.~V.~Gamayun}
\affiliation{Physics Department, Lancaster University, Lancaster, LA1 4YB, UK}
\affiliation{Bogolyubov Institute for Theoretical Physics, 03680, Kiev, Ukraine}

\author{E.~V.~Gorbar}
\affiliation{Department of Physics, Taras Shevchenko National Kiev University, 03022, Kiev, Ukraine}
\affiliation{Bogolyubov Institute for Theoretical Physics, 03680, Kiev, Ukraine}

\author{V.~P.~Gusynin}
\affiliation{Bogolyubov Institute for Theoretical Physics, 03680, Kiev, Ukraine}

\begin{abstract}
It is known that the symmetric (massless) state of the Nambu--Jona-Lasinio model
in 2+1 dimensions in a magnetic field $B$ is not the ground state of the system at
zero temperature due to the presence of a negative, linear in $|\sigma+i\pi|$, term in the effective
potential for the composite fields $\sigma\sim\bar{\psi}\psi$ and $\pi\sim\bar{\psi}i\gamma^5\psi$,
while the quadratic term is always positive (a tachyon is absent).
We find that finite temperature is a necessary ingredient for the tachyonic instability
of the symmetric state to occur. Utilizing the Schwinger--Keldysh real-time formalism we
calculate the dispersion relations for the fluctuation modes of the composite fields
$\sigma$ and $\pi$. We demonstrate the presence of the tachyonic instability of the
symmetric state for coupling constant that exceeds a certain critical value which vanishes
as temperature tends to zero in accordance with the phenomenon of magnetic catalysis.
\end{abstract}
\maketitle

\section{Introduction}

For many years relativistic quantum field models in (2+1) dimensions have attracted a significant
interest both due to their sophisticated dynamics and the fact that they describe
long wavelength excitations in several planar condensed matter systems \cite{Jackiw},
among them graphene \cite{Semenoff-graphene}, the d-wave state of high $T_{c}$ superconductors \cite{d-wave},
topological insulators \cite{top-insulators} and optical lattices \cite{opt-lattice}.
Recently there has been a surge of activity in this area connected with the experimental
discovery of graphene \cite{Novoselov} whose quasiparticle excitations
are described by the massless Dirac equation in (2+1) dimensions that leads to many unusual
electronic properties of this material and opens new perspectives for electronic
devices (see, review papers \cite{graphene-review}). Lattice effects necessarily
produce local interactions for quasiparticles in graphene \cite{Alicea} and, thus, one naturally
comes at the gauged Nambu--Jona-Lasinio (NJL) model in 2+1 dimensions.

Historically the Nambu--Jona-Lasinio model \cite{NJL} was the first model in which
the mass generation and dynamical symmetry breaking (DSB) were considered in elementary particle physics
and quantum field theory. At present NJL-type models have a significant
practical value, for example, the NJL model provides a successful effective theory
of low energy Quantum Chromodynamics \cite{Kleinert,Volkov,Hatsuda}.
Dynamical symmetry breaking occurs in the NJL model only in supercritical regime when its
coupling constant $G$ exceeds a critical value $G_c$. This is different from
the Bardeen--Cooper--Schriffer (BCS) theory where a gap in
quasiparticle spectrum is generated for any value of coupling constant. The physical
reason for zero value of the critical coupling constant is connected with the presence of
the Fermi surface in the BCS theory. According to the renormalization-group studies \cite{Shankar},
the renormalization-group scaling takes place only in the direction perpendicular to
the Fermi surface that lowers effectively the space-time dimension by two units to
a (1+1)-dimensional theory, where as is well known, symmetry breaking occurs for
arbitrary weak attraction between fermions.

Since dynamical symmetry breaking in (3+1) and (2+1)-dimensional theories requires strong
coupling ($g_c \ge 1$) it makes the quantative study of DSB a difficult problem. Therefore,
it is very interesting to consider field-theoretical models where DSB takes place in the regime of weak coupling
($g_c \approx 0$). The DSB in a magnetic field \cite{catalysis1,catalysis2} (magnetic catalysis)
gives the corresponding example (see also Refs.\cite{catalysis3} and a short review Ref.\cite{UFG}).
The essence of the magnetic catalysis phenomenon is that the dynamics
of the electrons in a magnetic field, $B$, corresponds effectively to a theory with spatial dimension
reduced by two units (note a close similarity with the role of the Fermi surface in the BCS theory)
if their energy is much less than the Landau gap $\sqrt{|eB|}$. The zero-energy Landau level
has a finite density of states and this is a key ingredient of magnetic catalysis
which plays, in fact, the role of the Fermi surface.

The magnetic catalysis is an universal phenomenon and its main features are model independent
\cite{catalysis1,Semenoff}. It was studied, besides a (2+1)-dimensional NJL-type model, in the NJL$_{3+1}$
 model \cite{catalysis2}, quantum electrodynamics \cite{catalysis-QED}, and quantum chromodynamics
\cite{QCD}. The universality of this phenomenon is confirmed by applying holographic techniques
which have proven to be a powerful analytic tool in studying the qualitative properties of
strongly interacting physical systems such as interacting quark gluon plasma, graphene,
superconductivity, and superfluidity \cite{holography}.

In the theory of superconductivity the normal state of a superconductor is unstable at sufficiently
low temperature with respect to the transition to a superconducting state. This instability is
signaled by a pole in the scattering amplitude of the electrons with opposite momenta and is known
as the Cooper instability \cite{Schrieffer-book}. This instability is resolved  in
the superconducting state through the formation of a condensate of Cooper pairs \cite{Cooper}
that opens a gap in the electron quasiparticle spectrum.

The instability of the normal state of a quantum statistical or field system has a
precursor in the corresponding one particle problem which is
known as the fall-into-the-center phenomenon. For example, in the study of dynamical chiral
symmetry breaking in strongly coupled QED \cite{Rivista}, the corresponding one particle problem
is formulated as the Dirac equation for the electron in the field of the Coulomb center and
the precursor of the normal state instability in QED corresponds to the supercritical
charge problem when the lowest in energy bound state dives into the lower continuum. Then an
electron-positron pair is spontaneously created from vacuum with the electron shielding the
supercritical charge and positron emitted to infinity (described by a resonance state) \cite{Zeldovich,Greiner}.

It is interesting to see what is a precursor of the magnetic catalysis phenomenon in
quantum field theories and what are its characteristics. Recently the corresponding study was
performed in the case of graphene in Ref.\cite{3G}, where the Dirac equation for the electron
in the field of the Coulomb center in a magnetic field was considered and it was shown that, as suggested by
the magnetic catalysis phenomenon, indeed any charge in the gapless theory is supercritical. However,
no resonance state was found that is related to the fact that charged particles cannot propagate
freely to infinity in a constant magnetic field in two dimensions. Still it was found that the low
energy bound state crosses the level of filled states that suggests that the normal state of
the system in a magnetic field should suffer from a tachyonic instability
(i.e, an analog of the Cooper instability in the theory of superconductivity should exist).

In the present paper, we directly address the problem of instability of the symmetric state
of quantum field theories with attraction between fermions and antifermions in a magnetic field
in the framework of the NJL$_{2+1}$ model. The model is described in Sec.~\ref{secII}. The analysis of
the effective potential indicates the necessity of finite temperature for the tachyonic instability to
be present. In Sec.~\ref{secIII}, using the Schwinger--Keldysh real-time formalism, we calculate the
dispersion relations for composite fields in the LLL approximation
and for sufficiently low temperature find a tachyonic instability. The contribution of higher
Landau levels to the dispersion relations for composite fields is considered in Sec.~\ref{sec4}.
The main results are summarized in Conclusion.

\section{Model and effective potential}
\label{secII}

The NJL action in (2+1) dimensions in a magnetic field reads
\be
S =  \int d^3x  \left[\,\bar{\psi}i\gamma^{\mu}D_{\mu}\psi + \frac{G_0}{2}\left[\,
(\bar{\psi}(x)\psi(x))^2 + (\bar{\psi}(x)i\gamma^5\psi(x))^2\right]\,\right]\,,
\label{NJL-action}
\ee
where $D_{\mu}=\partial_{\mu}+ieA_{\mu}$ with the vector potential $A_{\mu}=(0,Bx,0)$
that describes a constant magnetic field in the Landau gauge. We use four-components spinors
corresponding to a reducible representation of the Dirac algebra like in Ref.\cite{catalysis1}.
According to magnetic catalysis,
we expect that the symmetric state of model (\ref{NJL-action}) is unstable for any $G_0 > 0$.
In order to see this, we calculate the effective potential for the composite fields
$\sigma\sim\bar{\psi}\psi$ and $\pi\sim\bar{\psi}i\gamma^5\psi$.

Using the Hubbard--Stratonovich method of auxiliary fields, model (\ref{NJL-action})
can be equivalently rewritten as follows:
\be
S_{aux}=\int d^3x  \left[\,\bar{\psi}(i\gamma^{\mu}D_{\mu}-\sigma-i\pi\gamma^5)\psi -
\frac{\sigma^2+\pi^2}{2G_0}\,\right]\,.
\label{auxiliary}
\ee
Assuming that $\sigma=const$ and $\pi=const$, the effective potential for composite
fields $\sigma$ and $\pi$ was found in the second paper in \cite{catalysis1} (for more details
of calculation see Ref.\cite{Incera}). The following
propagator for fermions with mass $m=\langle \sigma \rangle$ was used in the derivation:
\be
G(x,x')=e^{i\Phi(\mathbf{x},\mathbf{x}')}\tilde{G}(x-x'),
\ee
where the Schwinger phase \cite{Schwinger} is separated from the translation invariant
part $\tilde{G}(x-x')$. The translation invariant part of the propagator can be expanded
over the Landau levels (compare with Ref. \cite{catalysis1}) and in the mixed
$\omega, \mathbf{r}$ representation it has the form
\ba
\tilde{G}(\omega; \mathbf{r})&=&\frac{i}{2\pi l^{2}}
\exp\left(-\frac{\mathbf{r}^{2}}{4l^{2}}\right)\sum\limits_{n=0}^{\infty}
\frac{1}{\omega^{2}-E^{2}_{n}+i\epsilon}\left\{(\gamma_{0}\omega+m)
\left[{\cal P}_{-}L_{n}\left(\frac{\mathbf{r}^{2}}{2l^{2}}\right)+{\cal P}_{+}L_{n-1}
\left(\frac{\mathbf{r}^{2}}{2l^{2}}\right)\right]\right.\nonumber\\
&-&\left.\frac{i}{l^{2}}\pmb{\gamma}\mathbf{r}L^{1}_{n-1}
\left(\frac{\mathbf{r}^{2}}{2l^{2}}\right)\right\},
\label{free-propagator}
\ea
where ${\cal P}_{\pm}=(1\pm i\,{\rm sgn}(eB)\gamma_{1}\gamma_{2})/2$, $E_{n}=
\sqrt{m^2+2|eB|n}$ are the Landau levels energies, $l=1/\sqrt{|eB|}$ is the magnetic length,
functions $L^{\alpha}_{n}(x)$ are the generalized Laguerre polynomials,
and by definition, $L_n(x)= L^{0}_n(x)$, $L^{\alpha}_{-1}(x)\equiv 0$. Further, according to
\cite{catalysis1,catalysis2,catalysis-QED}, the lowest Landau level (LLL) contribution is
responsible for zero value of the critical coupling constant. Since we are interested in
the instability of the normal state of the model in the weak coupling regime, it is clear
that only the dynamics in the LLL can produce this instability.
Eq.(\ref{free-propagator}) implies that the LLL fermion propagator in momentum space is given by
\be
\tilde{G}_{LLL}(\omega; \mathbf{p})=i\,e^{-\mathbf{p}^{2}l^{2}}
\frac{\mathcal{P}_-}{\omega^{2}-m^2+i\epsilon}\,.
\ee
The effective potential for composite fields $\sigma$ and $\pi$ at zero temperature
and zero chemical potential in the model under consideration was calculated in Ref. \cite{catalysis1},
\be
V(\rho)=\frac{1}{\pi}\left[\,\frac{\Lambda}{2\sqrt{\pi}}\left(\frac{\sqrt{\pi}}{g}-1\right)
\rho^2-\frac{\sqrt{2}}{l^3}\zeta\left(-\frac{1}{2};\frac{(\rho l)^2}{2}+1\right)
-\frac{\rho}{2l^2}\,\right]\,,
\label{effective-potential}
\ee
where $\rho=\sqrt{\sigma^2+\pi^2}$, $g=G_0\Lambda/\pi$, $\Lambda$ is the UV cut-off,
and $\zeta(s,q)$ is the generalized Riemann zeta function. For $\rho \to 0$, at weak
coupling $g \ll \sqrt{\pi}$ we have
\be
V(\rho)\approx\frac{\rho^2}{2G_0}-\frac{\rho}{2\pi l^2}\,.
\label{effective-potential-origin}
\ee

The presence of negative linear term $\rho/(2\pi l^2)$ clearly indicates that the true
minimum of the effective potential corresponds to a state with broken symmetry. However,
the second derivative of the effective potential with respect to $\rho$ is always positive, hence the
tachyon is absent. This situation is rather unusual and the reason for the existence of
the linear term was explained in \cite{catalysis1}. Since the gap equation is given by
$\partial V(\rho)/\partial\rho=0$,
the effective potential can be reconstructed up a constant by integrating the gap equation.
It suffices to consider only the field $\sigma$ (the dependence of the effective potential
on $\pi$ can be easily restored using the chiral symmetry).  Further, the gap equation in
the model under consideration equals
\be
\sigma=G_0\langle 0|\bar{\psi}\psi|0 \rangle\,.
\label{gap-equation}
\ee
The point crucial for the existence of the linear term in the effective potential is that
the chiral condensate $\langle 0|\bar{\psi}\psi|0 \rangle $ does not vanish as
$\sigma \to 0$ even in the free noninteracting theory. It suffices to keep only the LLL
contribution. Then we have
\be
\langle 0|\bar{\psi}\psi |0 \rangle =-i\lim_{\sigma \to 0}\frac{4\sigma}{(2\pi)^3}
\int d\omega d^2\mathbf{p}\,\frac{e^{-\mathbf{p}^2l^{2}}}{\omega^2-\sigma^2}=-\frac{1}{2\pi l^2}\,.
\label{condensate}
\ee
Integrating it, we find the linear term in the effective potential (\ref{effective-potential-origin}).
Since fluctuations are described by quadratic terms in the effective action, the linear
term present in the effective potential does not influence fluctuations. Consequently,
a tachyonic instability is {\sl absent} in the spectrum of the model at zero temperature.

The situation changes crucially at finite temperature. At $T \ne 0$ the effective potential
was calculated in Refs. \cite{catalysis1,Incera}. For small $\rho/T \ll 1$,
at weak coupling it is given by
\be
V_T(\rho)=\left(\,\frac{1}{G_0}-\frac{1}{G_{c}(T,B)}\,\right)\frac{\rho^2}{2},\quad G_{c}(T,B)=4\pi Tl^2\,.
\label{effective-potential-origin-T}
\ee
Clearly, for $G_{0}>G_{c}(T,B)$, we have an instability of the conventional (tachyonic) type.
The critical coupling $G_{c}(T,B)$ tends to zero as $T\to0$, and the symmetry broken ground state occurs
at arbitrary small attractive interaction in accordance with the phenomenon of magnetic catalysis.
The absence of the linear term at finite temperature is consistent with the absence of
chiral condensate in the free theory at $T \ne 0$,
\be
\langle 0|\bar{\psi}\psi |0 \rangle_T =-\lim_{\sigma \to 0}\frac{4\sigma T}{(2\pi)^2}
\sum_{n=-\infty}^{+\infty}\int d^2\mathbf{p}\,\frac{e^{-\mathbf{p}^2l^{2}}}{(\pi T(2n+1))^2+\sigma^2}=
-\lim_{\sigma \to 0}\frac{1}{2\pi l^{2}}\tanh\frac{\sigma}{2T}
=0\,.
\label{condensate-T}
\ee
This result suggests that in order to find a tachyonic instability, we should study
quadratic fluctuations of the model at finite temperature. To do this, we will calculate
in the next section the dispersion relations for composite fields $\sigma$ and $\pi$ at
finite temperature in the LLL approximation and analyze them.

\section{Tachyonic instability for composite fields in the real-time formalism
in the LLL approximation}
\label{secIII}

The analysis in the previous section shows that in the model under consideration
a tachyonic instability can appear only at finite temperature. Since instability
is an inherently dynamical process and the Matsubara imaginary time formalism is
mainly used for the study of theories at thermodynamical equilibrium, in order to
analyze the tachyonic instability we will utilize the Schwinger--Keldysh real-time
formalism \cite{Schwinger1,Keldysh} (for a review, see \cite{real-time-review}).
The action in the real-time formalism contains integrals over positive time branch
$t_+$ and negative one $t_-$. Then action (\ref{auxiliary}) transforms into
\be
{\rm S} =  \int_{p} d^3x\,\, \bar{\psi}(i\gamma^{\mu}D_{\mu}-\sigma-i\pi\gamma^5)
\psi - \int_{p} d^3x\,\,\frac{\sigma^2 + \pi^2}{2G_0}\,,
\label{action-real-time}
\ee
where the time integration proceeds along the closed path time contour \cite{real-time-review}
\be
\int_{p} dt=\int_{-\infty}^{+\infty} dt_+ + \int_{+\infty}^{-\infty} dt_- =
\int_{-\infty}^{+\infty} dt_+ - \int_{-\infty}^{+\infty} dt_-\,.
\label{closed-path-time-contour}
\ee
Since the fields $\sigma_{\pm}$ and $\pi_{\pm}$ are defined on the positive and negative
time parts of the contour, in what follows it is convenient to consider their linear combinations
\be
\sigma_{\rm c,\, \Delta} = \frac{\sigma_{+}\pm \sigma_{-}}{2},\,\,\,\,\,\,\,
\pi_{\rm c,\, \Delta} = \frac{\pi_{+}\pm \pi_{-}}{2}\,.
\ee
Integrating over fermions in the functional integral, we find the following
effective action for the composite fields:
\be
{\rm S}_{\rm eff} = -\int_{p}d^3x\,  \frac{\sigma^{2}+\pi^{2}}{2G_0}\,
-i{\rm Tr}\ln\left[G^{-1}\right]\,,
\label{effective-action-real-time}
\ee
where $G^{-1}=-i(i\gamma^{\mu}D_{\mu}-\sigma-i\pi\gamma^5)\delta_{p}(x-y)$
and $\delta_{p}(x-y)$ is a contour $\delta$-function,
the trace ${\rm Tr}$ in Eq.(\ref{effective-action-real-time}) is taken in the functional sense.
The equations of motion for composite fields follow from this effective action and
are given by (the physically sensible case corresponds to $\sigma_{\Delta}=\pi_{\Delta}=0$)
\cite{real-time-review,Fu}
\ba
\frac{2\sigma_{\rm c}(x)}{G_0} &=&-i\frac{\delta {\rm Tr}\ln\left[G^{-1}
 \right]}{\delta \sigma_{\Delta}(x)} \Big|_{ \sigma_{\Delta}=\pi_{\Delta}=0}=-i{\rm Tr}\left[
 G\frac{\delta G^{-1}}{\delta \sigma_{\Delta}(x)}\right]\Big|_{ \sigma_{\Delta}=\pi_{\Delta}=0}\,,\\
\label{equation-motion-1}
\frac{2\pi_{\rm c}(x)}{G_0} &=&-i\frac{\delta {\rm Tr}\ln\left[G^{-1}\right]}{\delta \pi_{\Delta}(x)}
\Big|_{ \sigma_{\Delta}=\pi_{\Delta}=0}=-i{\rm Tr}\left[
 G\frac{\delta G^{-1}}{\delta \pi_{\Delta}(x)}\right]\Big|_{ \sigma_{\Delta}=\pi_{\Delta}=0}\,.
\label{equation-motion-2}
\ea
Here $G(x,y)$ is the two-point correlation function defined as
\ba
G(x,y)={\rm Tr}(T_{p}(\psi(x)\bar\psi(y))\hat\rho)\equiv\langle T_{p}(\psi(x)\bar\psi(y)\rangle,
\ea
$\hat\rho$ is the thermal density matrix and $T_{p}$ is the time-ordering operator along a complex path $p$.
Since $x,y$ can take values on either positive or negative  time branches, it is convenient to represent
$G(x,y)$ as $2\times2$ matrix:
\ba
G(x,y)=\left(\begin{array}{cc}G_{++}&G_{+-}\\
G_{-+}&G_{--}\end{array}\right)=\left(\begin{array}{cc}\langle T(\psi(x)\bar\psi(y)\rangle
&-\langle \bar\psi(y)\psi(x)\rangle\\
\langle\psi(x)\bar\psi(y)\rangle&\langle\tilde{T}\psi(x)
\bar\psi(y)\rangle\end{array}\right).
\ea
where $T$ and $\tilde{T}$ are the usual time-ordering operator and  anti-time-ordering operators,
respectively.  Note the identity $G_{++}+G_{--}=G_{+-}+G_{-+}$ which follows from the identity
for the step functions $\theta(x-y)+\theta(y-x)=1$.

Since we are interested in physical excitations, we will consider time dependent solutions
of the above equations which deviate weakly from constant values $\sigma_{\rm c}(x) =
\tilde{\sigma}(x)+\bar{\sigma}$ and $\pi_{\rm c}(x) = \tilde{\pi}(x)$, where $\bar{\sigma}=const$.
Then we obtain
\be
\frac{2(\tilde{\sigma}(x)+\bar{\sigma})}{G_0} =-i{\rm Tr}\left[
 G\frac{\delta G^{-1}}{\delta \sigma_{\Delta}(x)}\right] \Big|_{\sigma_{\Delta}=\pi_{\Delta}=0,\,
\sigma_c=\bar{\sigma}}+i
\int_{p} d^3y\, {\rm Tr}\left[G\frac{\delta G^{-1}}{\delta \sigma_{\Delta}(x)}
G\frac{\delta G^{-1}}{\delta \sigma_{c}(y)} \right]
 \Big|_{ \sigma_{\Delta}=\pi_{\Delta}=0,\,
\sigma_{\rm c}=\bar{\sigma}}\,\tilde{\sigma}(y)\,,\\
\label{equation-motion-1-1}
\ee
\be
\frac{2\tilde{\pi}(x)}{G_0} =-i{\rm Tr}\left[
 G\frac{\delta G^{-1}}{\delta \sigma_{\Delta}(x)}\right] \Big|_{ \sigma_{\Delta}
 =\pi_{\Delta}=0,\,\sigma_c=\bar{\sigma}}+i
\int_{p} d^3y\, {\rm Tr}\left[G\frac{\delta G^{-1}}{\delta \pi_{\Delta}(x)}
G\frac{\delta G^{-1}}{\delta \pi_{c}(y)} \right] \Big|_{ \sigma_{\Delta}=\pi_{\Delta}=0,\,\sigma_{\rm c}
=\bar{\sigma}}\tilde{\pi}(y)\,.
\label{equation-motion-2-2}
\ee
In the single time representation $\delta_{p}(x-y)=\tau_{3}\delta(x-y)$, where $\tau_{3}$ is the third Pauli
matrix, and the variational derivatives and functional traces are calculated according to the rules
\be
\frac{\delta G^{-1}(x,y)}{\delta\sigma_{\Delta}(z)}=i\frac{\delta\sigma(x)}{\delta\sigma_{\Delta}(z)}\,
\delta_{p}(x-y)=i\delta_{p}(x-y)\delta_{p}(x-z)=i\delta(x-y)\delta(x-z),
\ee
\ba
 {\rm Tr}\left[G\frac{\delta G^{-1}}{\delta \sigma_{\Delta}(x)}\right]
&=&\int_{p}d^{3}ud^{3}v\,{\rm tr}\left[G(u,v)\frac{\delta G^{-1}(v,u)}{\delta \sigma_{\Delta}(x)}\right]
\nonumber\\
&=&\int d^{3}ud^{3}v\,{\rm tr}\left[\tau_{3}G(u,v)\tau_{3}\frac{\delta G^{-1}(v,u)}
{\delta \sigma_{\Delta}(x)}\right]
=\int d^{3}ud^{3}v\,{\rm tr}\left[G(u,v)\frac{\delta G^{-1}(v,u)}
{\delta \sigma_{\Delta}(x)}\right].
\ea
To calculate the right-hand sides of Eqs.(\ref{equation-motion-1-1}) and (\ref{equation-motion-2-2}),
we should determine the fermion Green`s function in the real-time formalism. As we discussed
in the previous section, for our purposes it suffices to use the LLL approximation.
The LLL Green`s function in the real-time formalism equals
\be
G_{LLL}(x,x') = \mathcal{P}_- K(\mathbf{x},\mathbf{x}')\left(
\begin{array}{cc}
G^{++}(t-t') & G^{+-}(t-t') \\
G^{-+}(t-t') & G^{--}(t-t') \\
\end{array}
\right)\,,
\label{real-time-LLL-propagator}
\ee
where
\be
K(\mathbf{x},\mathbf{x}')=\frac{1}{2\pi l^2}
\exp\left[-\frac{(\mathbf{x}-\mathbf{x}')^2}{4l^2}+i\Phi(\mathbf{x},\mathbf{x}')\right]
\label{phase-factor}
\ee
is the space dependent part of the LLL fermion propagator and
\be
G^{++}(\omega) = i\frac{\gamma_0\omega+\bar{\sigma}}{\omega^2-\bar{\sigma}^2+i\epsilon}
- 2\pi  (\gamma^0\omega+\bar{\sigma})n_F(\bar{\sigma})\delta(\omega^2-\bar{\sigma}^2)\,,
\ee
\be
G^{--}(\omega) = -i\frac{\gamma_0\omega+\bar{\sigma}}{\omega^2-\bar{\sigma}^2-i\epsilon}- 2\pi
(\gamma^0\omega+\bar{\sigma})n_F(\bar{\sigma})\delta(\omega^2-\bar{\sigma}^2)\,,
\ee
\be
G^{+-}(\omega) = -2\pi \left(\gamma^{0}\omega+\bar{\sigma}\right)n_F(\omega){\rm sgn}(\omega)
\delta(\omega^{2}-\bar{\sigma}^{2})\,,
\ee
\be
G^{-+}(\omega) = 2\pi \left(\gamma^{0}\omega+\bar{\sigma}\right)n_F(-\omega){\rm sgn}(\omega)
\delta(\omega^{2}-\bar{\sigma}^{2})
\ee
are the Fourier transforms of  $G^{ij}(t),i,j=\pm$. Here $n_F(\bar{\sigma})
=(\exp(\bar{\sigma}/T)+1)^{-1}$ is the Fermi--Dirac distribution function.

Further, it is convenient to perform the unitary Keldysh transformation \cite{Keldysh,real-time-review}
\ba
G\,\,\, \to\,\,\, U^{\dagger}G U = \left(\begin{array}{cc}
 0 & G_a \\
 G_r & G_c \\
\end{array}\right)\,, \quad U=\frac{1}{\sqrt{2}}\left(\begin{array}{cc}
 1 & 1 \\
 -1 & 1 \\ \end{array}\right),
 \label{Keldysh-transformation}
\ea
where
\ba
G_a(\omega) &=& \frac{1}{2} \left(G^{++}-G^{--}+G^{+-}-G^{-+}\right)=G^{++}-G^{-+} = i\frac{\gamma^0\omega
+\bar{\sigma}}{\omega^2 - \bar{\sigma}^2 - i\epsilon{\rm sgn}\omega}\,,\\
G_r(\omega) &=& \frac{1}{2} \left(G^{++}-G^{--}-G^{+-}+G^{-+}\right)=G^{++}-C^{+-}= i\frac{\gamma^0\omega
+\bar{\sigma}}{\omega^2 - \bar{\sigma}^2 + i\epsilon{\rm sgn}\omega}\,,\\
G_c(\omega) &=& G^{++}+G^{--} = G^{+-}+G^{-+}=2\pi\tanh\frac{\bar{\sigma}}{2T}\,(\gamma^0 \omega+\bar{\sigma})\,
\delta(\omega^2 -\bar{\sigma}^2)
\ea
are the advanced, retarded, and correlation functions. For time dependent and spatially
homogeneous modes $\tilde{\sigma}(t)$ and $\tilde{\pi}(t)$, Eqs.(\ref{equation-motion-1-1})
and (\ref{equation-motion-2-2}) imply the following equations:
\be
\tilde{\sigma}(t) = {G_0}\int dt'\,\Pi^{\sigma}(t-t')\,\tilde{\sigma}(t')\,,
\label{first-equation}
\ee
\be
\tilde{\pi}(t) = {G_0}\int dt'\,\Pi^\pi(t-t')\tilde{\pi}(t')\,,
\label{second-equation}
\ee
where
\be
\Pi^{\sigma}(t-t') =\frac{-i}{4\pi l^2}{\rm tr}\left[G_r(t-t')G_c(t'-t)+G_c(t-t')G_a(t'-t)
\right] = 0\,,
\label{first-coefficient}
\ee
\be
\Pi^\pi(t-t') =\frac{-i}{4\pi l^2}{\rm tr}\left[
i\gamma^5 G_r(t-t')i\gamma^5 G_c(t'-t)+i\gamma^5 G_c(t-t') i\gamma^5 G_a(t'-t)
\right] =  \frac{2{\bar \sigma}}{\pi l^{2}}\tanh\frac{\bar{\sigma}}{2T}\int \frac{d\Omega}{2\pi}
\frac{e^{-i\Omega(t-t')}}{4\bar{\sigma}^2-\Omega^2}\,.
\label{second-coefficient}
\ee
Hence Eq.(\ref{first-equation}) gives $\tilde{\sigma}(t)=0$.
Note that the equality $\Pi^{\sigma}(t)=0$ is due to
the LLL approximation used in this section. On the other hand, $\Pi^\pi(t)\neq0$ in the same
approximation. In the next section we obtain expressions for $\Pi^{\sigma,\pi}(t)$
where all Landau levels are taken into account.

For the Fourier transform $\tilde{\pi}(\Omega)$, we find
\be
\left(
\frac{2\pi l^2}{G_0}
-\frac{{4\bar \sigma}\tanh\frac{\bar \sigma}{2T}}{4\bar{\sigma}^2-\Omega^2}
\right)\tilde{\pi}(\Omega) = 0
\ee
that implies
\be
\Omega^2 = 4\bar{\sigma} \left[{\bar \sigma} - \frac{G_0}{2\pi l^2}\tanh\frac{\bar \sigma}{2T}
\right]\,.
\ee
For $\bar{\sigma} \to 0$,
\be
\Omega^2=4\bar{\sigma}^2 \left[1- \frac{G_0}{4\pi Tl^2}\right]\,.
\label{tachyon}
\ee
Obviously, for $T$ less than the critical value
\begin{equation}
T_{c}=\frac{G_0}{4\pi l^2}\,,
\label{critical-constant}
\end{equation}
we have a tachyon.
This result is perfectly consistent with the effective potential at finite temperature
(\ref{effective-potential-origin-T}) whose symmetric and symmetry broken phases are separated
 by the curve
\begin{equation}
\frac{1}{G_0}-\frac{1}{4\pi l^2T}=0,
\label{crit-line}
\end{equation}
that leads to the critical temperature (\ref{critical-constant}).
\vspace{3mm}

\section{The effective action for composite fields in the real-time formalism
beyond the LLL approximation}
\label{sec4}

In Sec. III, we calculated the correlators $\Pi^{\sigma}(t-t')$ and $\Pi^\pi(t-t')$
given by Eqs.(\ref{first-coefficient})
and (\ref{second-coefficient}) in the LLL approximation. In the present section,
we calculate these quantities taking into account the contribution of all Landau levels.
In addition, we determine the dependence of $\Pi^{\sigma}$ and $\Pi^\pi$ on spatial coordinates,
i.e. calculate $\Pi^{\sigma}(t-t',\mathbf{x}-\mathbf{x'})$ and $\Pi^\pi(t-t',\mathbf{x}-\mathbf{x'})$
(note that $\Pi^{\sigma}$ and $\Pi^\pi$ are translation invariant in spatial coordinates because
the Schwinger phases cancel out for a closed fermion loop with two vertices).

In the real time formalism the propagator in a magnetic field and at finite temperature
can be written in the form
\be
G(x,y)= K(\mathbf{x},\mathbf{y})\int\frac{d\omega}{2\pi} e^{-i(x^0-y^0)\omega}\sum\limits_{n=0}^{\infty}
D_n(\mathbf{x}-\mathbf{y},\omega)
\left(
\begin{array}{cc}
G_n^{++}(\omega) & G_n^{+-}(\omega) \\
G_n^{-+}(\omega) & G_n^{--}(\omega) \\
\end{array}\right),
\label{propagator-alllevels}
\ee
where the factor $K(\mathbf{x},\mathbf{y})$ is  given by Eq.(\ref{phase-factor}) and
\be
D_n (\mathbf{r},\omega) = (\gamma^0\omega+\bar\sigma)\left(\mathcal{P}_-L_n\left(\frac{\mathbf{r}^2}{2l^2}\right)
+\mathcal{P}_+L_{n-1}\left(\frac{\mathbf{r}^2}{2l^2}\right)
\right)- i  \frac{\bm{\gamma}\mathbf{r}}{ l^{2}}L^1_{n-1}\left(\frac{\mathbf{r}^2}{2l^2}\right).
\ee
Further,
\ba
G_n^{++}(\omega) &=& \frac{i}{\omega^2 -E_n^2+i0} -2\pi n_F(E_n) \delta(\omega^2-E^2_n),\\
G_n^{--}(\omega) &=& -\frac{i}{\omega^2 -E_n^2-i0} -2\pi n_F(E_n) \delta(\omega^2-E^2_n),
\ea
\be
G^{-+}_{n}(\omega) = 2\pi  n_{F}(-\omega){\rm sgn}(\omega)\delta(\omega^2-E_n^2),\,\,\,\,\,\,\,\,\,
G^{+-}_{n}(\omega) = -2\pi  n_{F}(\omega){\rm sgn}(\omega)\delta(\omega^2-E_n^2),
\ee
and $E_n=\sqrt{\bar\sigma^2+2|eB|n}$.

The correlators $\Pi^{\sigma}(x-x'), \Pi^\pi(x-x')$ ($x=(t,\mathbf{x})$) are defined by the expressions,
\ba
\Pi^{\sigma}(x-x')=-i{\rm Tr}\left[G(x,x')G(x',x)\right],\quad\quad
\Pi^\pi(x-x')=-i{\rm Tr}\left[i\gamma^{5}G(x,x')i\gamma^{5}G(x',x)\right],
\ea
and the trace ${\rm Tr}$ includes also the trace (${\rm tr}$) over Dirac indices.
Performing the Keldysh transformation the matrix in Eq.(\ref{propagator-alllevels}) takes the
form like in Eq.(\ref{Keldysh-transformation}) with
\ba
G_{r,a}(\omega,n)=\frac{i}{\omega^{2}-E_{n}^{2}\pm i\epsilon{\rm sgn}\omega}
=\frac{i}{(\omega\pm i\epsilon)^{2}-E_{n}^{2}},\,\,\,\,\,\,\,
G_{c}(\omega,n)=2\pi\tanh\frac{E_{n}}{2T}\,\delta(\omega^{2}-E_{n}^{2}).
\ea
Taking the Fourier transform of $\Pi^{\sigma}(t,\mathbf{x})$ we obtain
\ba
\Pi^{\sigma}(\Omega,\mathbf{k})&=&-i\int\frac{d\omega}{2\pi}\int d^{2}r\,e^{i\mathbf{k}\mathbf{r}}
{\rm tr}\left[G_{r}(\mathbf{r},\omega)
G_{c}(-\mathbf{r},\omega+\Omega)+G_{c}(\mathbf{r},\omega)
G_{a}(-\mathbf{r},\omega+\Omega)\right]\nonumber\\
&=&-i\sum\limits_{n,m=0}^{\infty}\int\frac{d\omega}{2\pi}\left[G_{r}(\omega,n)G_{c}(\omega+\Omega,m)
+G_{c}(\omega,n)G_{a}(\omega+\Omega,m)\right]\nonumber\\
&\times&\int \frac{d^{2}r}{(2\pi l^{2})^{2}}\,e^{i\mathbf{k}\mathbf{r}-\mathbf{r}^{2}/2l^{2}}
{\rm tr}\left[D_{n}(\mathbf{r},\omega)D_{m}(-\mathbf{r},\omega+\Omega)\right].
\ea
The space integral equals
\be
\int \frac{d^2r}{2\pi l^2} e^{i\mathbf{k}\mathbf{r}-\mathbf{r}^{2}/2l^{2}}{\rm tr}[
D_n (\mathbf{r},\omega)D_m (-\mathbf{r},\omega+\Omega)] =  2
s_{nm}(y)(\omega(\omega+\Omega)+\bar\sigma^2) - 4|eB|r_{nm}(y),
\ee
where
\ba
s_{nm}(y) &=& \int \frac{d^2 r}{4\pi l^2} e^{i\mathbf{k}\mathbf{r}-\mathbf{r}^{2}/2l^{2}}
\left(L_n\left(\frac{r^2}{2l^2}\right)L_m\left(\frac{r^2}{2l^2}\right)+L_{n-1}
\left(\frac{r^2}{2l^2}\right)L_{m-1}\left(\frac{r^2}{2l^2}\right)\right)
\label{snm-int}\nonumber\\
&=& \frac{(-1)^{n+m}}{2}e^{-y}\left( L_{m}^{n-m}(y)L_{n}^{m-n}(y)+L_{m-1}^{n-m}(y)L_{n-1}^{m-n}(y)\right)\,,
\quad y=\mathbf{k}^{2}l^{2}/2,\\
r_{nm}(y) &=&  \int \frac{d^2 r}{2\pi l^2}  \frac{\mathbf{r}^2}{2l^2}e^{i\mathbf{k}\mathbf{r}-\mathbf{r}^{2}/2l^{2}}
 L^1_{n-1}\left(\frac{r^2}{2l^2}\right)L^1_{m-1}\left(\frac{r^2}{2l^2}\right)
= (-1)^{m+n}e^{-y}mL_{m}^{n-m}(y)L_{n-1}^{m-n}(y)\,.
\label{rnm-int}
\ea
[for the evaluation of the integrals $s_{nm}(y),r_{nm}(y)$ see Appendix A in Ref.\cite{Pyatkoskiy}].
Therefore, we get
\ba
\Pi^{\sigma}(\Omega,k) &=& \frac{1}{\pi l^2} \int\limits_{-\infty}^{\infty}d\omega
\sum_{n,m=0}^{\infty}t_m\delta(\omega^2-E_m^2)\left[
\frac{((\omega-\Omega)\omega+\bar\sigma^2)s_{nm}(y)-2|eB| r_{nm}(y)}{(\omega-\Omega+i0)^2-E_n^2}
\right.\nonumber\\
&+&\left.
\frac{((\omega+\Omega)\omega+\bar\sigma^2)s_{nm}(y)-2|eB| r_{nm}(y)}{(\omega+\Omega-i0)^2-E_n^2}
\right]\nonumber\\
&=&\sum_{n,m=0}^{\infty}\left[\frac{t_m}{E_m}
\frac{E_m^2(E_m^2-E_n^2-\Omega^2)s_{nm}(y)-(E_n^2-E_m^2-\Omega^2)
(s_{nm}(y)\bar{\sigma}^2-2|eB|r_{nm}(y))}{\pi l^2((E_m+E_n)^2-\Omega^2)((E_m-E_n)^2-\Omega^2)}
+(m\leftrightarrow n)\right],
\ea
where $t_{m}=\tanh(E_{m}/2T)$. The calculation of
the correlator $\Pi^\pi$ results in the same expression except $\bar\sigma^{2}$ is replaced
by $-\bar\sigma^{2}$. Thus, we write
\ba
\Pi^{\sigma,\pi}(\Omega,k)=\sum_{n,m=0}^{\infty}\left[\frac{t_m}{E_m}
\frac{E_m^2(E_m^2-E_n^2-\Omega^2)s_{nm}(y)-(E_n^2-E_m^2-\Omega^2)
(\pm s_{nm}(y)\bar{\sigma}^2-2|eB|r_{nm}(y))}{\pi l^2((E_m+E_n)^2-\Omega^2)((E_m-E_n)^2-\Omega^2)}
+(m\leftrightarrow n)\right],
\label{polarizations-general}
\ea
where $\pm$ signs correspond to $\Pi^{\sigma}$ and  $\Pi^{\pi}$, respectively.

\begin{figure}
 \includegraphics[width=8cm]{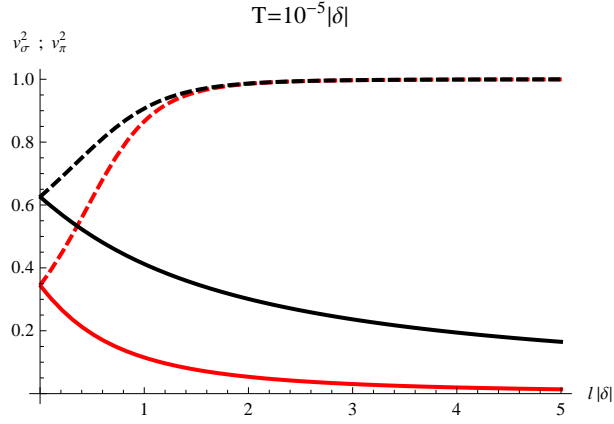}
\caption{(Color online) Velocities $v_{\pi}^2$ (red lines) and $v_{\sigma}^2$ (black lines) as
functions of $l|\delta|$ for temperature $T=10^{-5}|\delta|$. Solid (dashed) line corresponds to
$\delta>0$ ($\delta<0$).}
\label{velocities}
\end{figure}

\begin{figure}
 \includegraphics[width=8cm]{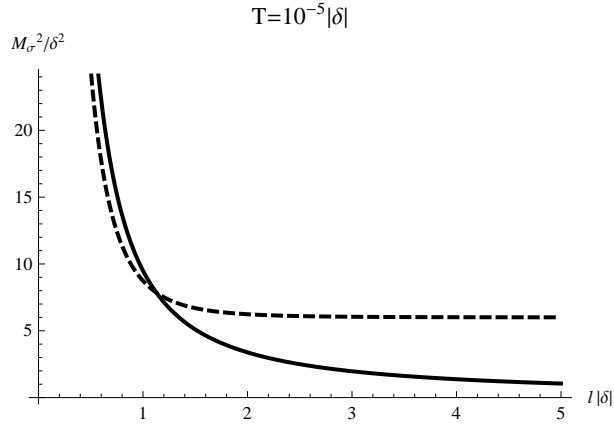}
\caption{$M^{2}_{\sigma}/\delta^{2}$ dependence on dimensionless magnetic field $l|\delta|$ for temperature
$T=10^{-5}|\delta|$. Solid (dashed) line corresponds to $\delta>0$ ($\delta<0$).}
\label{mass-sigma}
\end{figure}

To find the dispersion laws at small $\Omega$ and $|\mathbf{k}|$ it is convenient to evaluate
the sum over the Landau levels. This can be done explicitly if temperature is much lower
than the value of a magnetic field, $T\ll\sqrt{|eB|}$. The details of calculations are given
in Appendix. The dispersion relations  are given by the equations,
\be
-\frac{1}{G_{0}}+\Pi^{\sigma,\pi}(\Omega,k)=0.
\ee
For $\bar\sigma \ne 0$, the dispersion relations for $\tilde{\sigma}$ and $\tilde{\pi}$ modes at small
$\Omega l\ll 1$ and $y=\mathbf{k}^{2}l^{2}/2\ll 1$ take the form, respectively,
\ba
&&-\frac{1}{G_{0}}+\frac{V^{-}}{\pi^{3/2}l}+\Omega^{2}\frac{lQ^{-}}{4\pi^{3/2}}+y\left(
\frac{P^{-}}{\pi^{3/2}l}+\frac{\bar\sigma(1-\tanh({\bar\sigma}/{2T}))}{\pi}\right)=0,\\
&&-\frac{1}{G_{0}}+\frac{V^{+}}{\pi^{3/2}l}+\frac{\tanh{\bar\sigma}/{2T}-1}{2\pi l^{2}\bar\sigma}
+\Omega^{2}\left(\frac{lQ^{+}}{4\pi^{3/2}}+\frac{\tanh({\bar\sigma}/{2T})-1}{8\pi l^{2}\bar\sigma^{3}}\right)
+y\left(\frac{P^{+}}{\pi^{3/2}l}+\frac{1-\tanh({\bar\sigma}/{2T})}{2\pi l^{2}\bar\sigma}\right)=0,
\ea
where the quantities $V^{\pm},Q^{\pm},P^{\pm}$ are given by Eqs.(\ref{Vpm}) - Eqs.(\ref{Pminus}).
At the minimum of the effective potential the $\tilde{\pi}$ mode corresponds to
a Nambu-Goldstone boson and $\bar\sigma$ satisfies the gap equation,
\be
-\frac{1}{G_{0}}+\frac{V^{+}}{\pi^{3/2}l}+\frac{\tanh\frac{\bar\sigma}{2T}-1}{2\pi l^{2}\bar\sigma}=0.
\ee
The gap equation written in the form ($\zeta(s,v)$ is the generalized zeta function),
\be
- 2l\delta\bar\sigma+\frac{1}{l}\tanh\frac{\bar\sigma}{2T}+\sqrt{2}\bar\sigma
\zeta\left(\frac{1}{2},1+\frac{(\bar\sigma l)^{2}}{2}\right)=0,\quad\delta=\pi
\left(\frac{1}{G_{0}}-\frac{1}{G_{0c}}\right),\quad G_{0c}=\frac{2\pi^{3/2}}{3\Lambda},
\label{gap:eq}
\ee
is in agreement at $T=0$ with the one obtained in Ref.\cite{catalysis1}.
Fixing the intrinsic scale $\delta$ the gap equation determines $\bar\sigma$ as a function of  temperature
$T$ and  magnetic field $eB=1/l^{2}$ [We recall that in the used approximation $Tl\ll1$]. The critical
line separating symmetric and symmetry broken phases is obtained from Eq.(\ref{gap:eq})
when $\bar\sigma\to0$:
\be
\frac{1}{2T_{c}l}=2l\delta-\sqrt{2}\zeta\left(\frac{1}{2}\right),
\label{criticalline}
\ee
and in the weak coupling limit $G_{0}\ll G_{0c}$ it agrees with Eq.(\ref{crit-line}). The gap
equation (\ref{gap:eq}) was analyzed in Ref.\cite{catalysis1} at $T=0$ where three regions of
different behavior of $\bar\sigma$ as a function of a magnetic field were revealed. In the near
critical region $G_{0}\simeq G_{0c}$, where $|\delta| l\ll1$,  the dependence on the ultraviolet cutoff
$\Lambda$ disappears and we find $\bar\sigma l\simeq 0.45$. In other two regions, subcritical ($G_{0}<
G_{0c}$) and supercritical ($G_{0}>G_{0c}$), and for $|\delta| l\gg1$, the solution of the gap equation
behaves
\ba
\bar\sigma l&\simeq& \frac{1}{2\delta l}\ll1,\quad \delta>0,\\
\bar\sigma l&\simeq& |\delta|l\gg1,\quad \delta<0.
\ea
At finite temperature a nontrivial solution for $\bar\sigma$ in subcritical region ($\delta>0$)
exists for magnetic fields satisfying
\be
l\delta<\frac{1}{2\sqrt{2}}\left[\zeta(1/2)+\sqrt{\zeta^{2}(1/2)+\frac{2\delta}{T}}\right],
\ee
i.e., for magnetic fields exceeding some critical value. The dispersion laws  take the following form:
\be
\Omega^{2}=v_{\pi}^{2}\mathbf{k}^{2},\quad v_{\pi}^{2}=\frac{2\sqrt{2}(\bar\sigma l)^{2}}
{\sqrt{2}\tanh\frac{\bar\sigma}{2T}+(\bar\sigma l)^{3}\zeta\left(\frac{3}{2},1+\frac{(\bar\sigma l)^{2}}{2}\right)},
\label{dispersion-law-pi}
\ee
for the $\tilde{\pi}$ mode, and
\be
\Omega^{2}=M_{\sigma}^{2}+v_{\sigma}^{2}\mathbf{k}^{2},
\ee
\ba
 M_{\sigma}^{2}&=&\frac{8}{\bar\sigma l^{3}}\frac{\sqrt{2}\tanh\frac{\bar\sigma}{2T}+(\bar\sigma l)^{3}
 \zeta\left(\frac{3}{2},1+\frac{(\bar\sigma l)^{2}}{2}\right)}{2\zeta\left(\frac{3}{2},1+\frac{(\bar\sigma l)^{2}}
 {2}\right)-(\bar\sigma l)^{2}\zeta\left(\frac{5}{2},1+\frac{(\bar\sigma l)^{2}}{2}\right)},\\
 v_{\sigma}^{2}&=&4\frac{\sqrt{2}\bar\sigma l\tanh\frac{\bar\sigma}{2T}+(\bar\sigma l)^{2}\zeta\left(\frac{1}{2},
 1+\frac{(\bar\sigma l)^{2}}{2}\right)+\frac{(\bar\sigma l)^{4}}{2}\zeta\left(\frac{3}{2},1+
 \frac{(\bar\sigma l)^{2}}{2}\right)}
 {\zeta\left(\frac{3}{2},1+\frac{(\bar\sigma l)^{2}}{2}\right)-\frac{(\bar\sigma l)^{2}}{2}\zeta\left(\frac{5}{2},
 1+\frac{(\bar\sigma l)^{2}}{2}\right)},
\label{dispersion-law-sigma}
\ea
for the $\tilde{\sigma}$ mode, respectively. At zero temperature
Eqs.(\ref{dispersion-law-pi})-(\ref{dispersion-law-sigma}) coincide with those
obtained in Ref.[\onlinecite{catalysis1}]. One can check that the quantities $v_{\pi}^{2},v_{\sigma}^{2}$
are positive and  remain always less than $1$ (we set the velocity of light $c=1$). Their behavior for a chosen
value of a temperature ($T=10^{-5}|\delta|$) is shown in Fig.\ref{velocities} where the gap $\bar\sigma(T,l)$
is determined from Eq.(\ref{gap:eq}). The behavior of $M_{\sigma}^{2}/\delta^{2}$ as a function of the magnetic field
is shown in Fig.\ref{mass-sigma}. All dimensionful quantities in Figs.\ref{velocities},\ref{mass-sigma}
are measured in units of $|\delta|$.
Asymptotical behavior of the quantities $v_{\pi}^{2},v_{\sigma}^{2}$ in subcritical and supercritical
regions and for $|\delta l|\gg1$ is given by the expressions (for simplicity we take $T=0$),
\ba
v_{\pi}^{2}\simeq\left\{\begin{array}{cc}\frac{1}{2(\delta l)^{2}},\quad \delta>0,\\
1-\frac{1}{4(\delta l)^{4}},\quad \delta>0,\end{array}\right.\quad
v_{\sigma}^{2}\simeq\left\{\begin{array}{cc}\frac{2\sqrt{2}}{\zeta(3/2)\delta l},\quad \delta>0,\\
1-\frac{1}{6(\delta l)^{4}},\quad \delta>0,\end{array}\right.
\ea
and for the square mass $M^{2}_{\sigma}$,
\ba
M^{2}_{\sigma}\simeq\left\{\begin{array}{cc}\frac{8\sqrt{2}}{\zeta(3/2)}\frac{\delta}{l},\quad\delta>0,\\
6\delta^{2}\left(1+\frac{1}{2(\delta l)^{4}}\right)),\quad\delta<0.\end{array}\right.
\ea
These asymptotics should be compared with those obtained in Ref.[\onlinecite{catalysis1}].

\section{Conclusion}
\label{conclusion}

According to the magnetic catalysis phenomenon \cite{catalysis1}, an arbitrary weak attraction
between fermions and their antiparticles leads to chiral symmetry breaking and gap generation
in (3+1) and (2+1)-dimensional theories in a magnetic field. Consequently, the normal state
of these theories should be unstable in a magnetic field even in the weak coupling regime.
It is worth noting that the instability of the normal state in a magnetic field is
qualitatively different for theories in (3+1) and (2+1) dimensions. Since constant magnetic
field effectively reduces \cite{catalysis1,catalysis2} the spacetime dimension by two
units for fermions in the infrared region, (3+1)-dimensional theories are reduced to
effective (1+1)-dimensional theories, where bound states are easily formed in the weak
coupling regime and resonance states describing emitted antiparticles propagating to
infinity are realized in the standard way in the corresponding quantum mechanical one
particle problems.

As noted in Introduction, the situation is different in (2+1)-dimensional theories.
The dimensional reduction in a magnetic field means that the corresponding effective
theories are (0+1)-dimensional ones. Consequently, no emission to infinity is possible.
This conclusion is explicitly confirmed by the study of the (2+1)-dimensional Dirac
equation for the electrons in the field of the Coulomb center in graphene in a magnetic
field performed in Ref. \cite{3G} where no resonance state was found.

In the present paper, in order to study the normal state instability connected with
the magnetic catalysis phenomenon in a (2+1)-dimensional theory, we considered the
weakly coupled NJL$_{2+1}$ model in a magnetic field at finite temperature.
The choice of the model was made basically from the requirement of the simplicity
of analysis. Certainly, the generalization to the case of long range gauge models
would be of significant interest.

Using the Hubbard--Stratonovich method of auxiliary fields, we sought for tachyonic
excitations in the normal state of the NJL$_{2+1}$ model in a magnetic field at
finite temperature. We would like to note that the consideration of the theory at
finite temperature is a necessary feature of our analysis. As discussed in Sec.II,
although the symmetric state of the effective potential is unstable, its quadratic
form of fluctuations that follows from (\ref{effective-potential-origin}) is positive definite,
hence tachyonic excitations are absent. The situation changes at finite temperature,
where the effective potential (\ref{effective-potential-origin-T}) has the instability
typical for a second order phase transition. Utilizing the Schwinger--Keldysh
real-time formalism, the dispersion relations for the composite
fields $\bar{\psi}\psi$ and $\bar{\psi}i\gamma^5\psi$ were calculated in the LLL
approximation in Sec.III, and for temperature less than a critical one a tachyonic excitation
in the normal ground state was found. Thus, although there is no resonance state
in the quantum mechanical one particle problem, the corresponding quantum field-theoretic problem in
a magnetic field does have a tachyonic excitation in the normal state for temperature
less a critical one.
The contribution of higher Landau levels into dispersion relations for the composite
fields $\sigma$ and $\pi$ was taken into account in Sec.IV.

\vspace{5mm}

\centerline{\bf Acknowledgements}
\vspace{3mm}

We are grateful to V.A. Miransky and I.A. Shovkovy for useful
discussions. This work  is supported partially by the SCOPES grant No. IZ73Z0\verb|_|128026
of Swiss NSF, the grant SIMTECH No. 246937 of the European FP7 program, the joint
Ukrainian-Russian SFFR-RFBR grant No. F40.2/108,
and by the Program of Fundamental Research of the Physics and Astronomy Division
of the NAS of Ukraine. V.P.G. acknowledges a collaborative grant from the
Swedish Institute. O.V.G. is grateful to the ERC grant No. 279738 NEDFOQ for
financial support.

\appendix
\section{Closed form for the correlators $\Pi^{\sigma,\pi}$}
To perform the summation over the Landau levels in Eq.(\ref{polarizations-general}) we assume
that $T\ll\sqrt{|eB|}$. Then we can set $t_{m}=1$ for all $m\ge1$ while keeping
$t_{0}=\tanh(\bar\sigma/2T)$, and expression (\ref{polarizations-general}) takes the form
\ba
\Pi^{\sigma,\pi}(\Omega,k) &=&
\frac{2\bar{\sigma}}{\pi l^2} \left(\tanh\frac{\bar{\sigma}}{2T}-1\right) \sum\limits_{n=0}^{\infty}
\frac{-2|eB|n(1\pm1)-\Omega^2(1\mp1)}{(E_n^2-(\bar{\sigma}+\Omega)^2)(E_n^2-(\bar{\sigma}
-\Omega)^2)}s_{n0}\nonumber\\
&+&\frac{1}{\pi l^2} \sum_{n,m=0}^{\infty}\frac{E_m+E_n}{(E_m+E_n)^2-\Omega^2}
\left(s_{nm}(y) -\frac{\pm s_{nm}(y)\bar{\sigma}^2-2|eB|r_{nm}(y)}{E_nE_m}\right).
\label{approx-Pi}
\ea
Thus, in the considered approximation, the temperature dependence is described by the terms
in the first line of the above equation.
To calculate the first sum over the Landau levels in Eq.(\ref{approx-Pi}) we use the representation
$1/a =\int_{0}^{\infty}dt e^{-at}$ valid for ${\rm Re}\,a>0$, and take into account that $s_{0n}(y)=
s_{n0}(y)=y^{n}e^{-y}/2n!$. The evaluation of the second sum in Eq.(\ref{approx-Pi}) is more involved.
First, we use the chain of transformations
\ba
\frac{E_m+E_n}{(E_n+E_m)^2-\Omega^2}\left(1,\frac{1}{E_nE_m}\right) &=&
\int\limits_{-\infty}^{\infty} \frac{d\omega}{\pi }\frac{(\omega(\omega+i\Omega),1)}
{((\omega+i\Omega)^2+E_n^2)(\omega^2+E_m^2)}\nonumber\\
&=& \frac{1}{\sqrt{\pi}}\int\limits_0^{\infty}\frac{dt_1dt_2}{\sqrt{t_1+t_2}}
e^{\frac{\Omega^2 t_1 t_2}{t_1+t_2} - t_1E_n^2-t_2E_m^2} \left(\frac{t_1+t_2+2t_1t_2
\Omega^2}{2(t_1+t_2)^2},1\right),
\ea
valid for $\Omega^{2}<E_{0}^{2}$. Then the sum
\be
S(t_{1},t_{2})=\sum\limits_{n,m=0}^{\infty}s_{nm}(y)e^{-t_{1}E_{n}^{2}-t_{2}E_{m}^{2}}
\ee
is evaluated using the integral representation (\ref{snm-int}) and the summation formula,
\be
 \sum\limits_{n=0}^{\infty}L_{n}^{\alpha}(z)x^{n}=(1-x)^{-(\alpha+1)}
\exp\left(\frac{x z}{x-1}\right),\quad |x|<1.
\ee
Finally, the space integral over $r$ in Eq.(\ref{snm-int}) gives
\ba
S(t_1,t_2) &=&\frac{1}{2}\coth(|eB|(t_1+t_2))\exp\left(-\bar{\sigma}^2(t_1+t_2)-
\frac{2y\sinh|eB|t_1\sinh|eB|t_2}{\sinh|eB|(t_1+t_2)}\right).
\ea
Similarly, for another sum we obtain ($r_{n0}(y)=r_{0m}(y)=0$),
\ba
R(t_1,t_2) \equiv\sum_{n,m=1}^{\infty}r_{nm}(y)e^{-t_1E_n^2-t_2E_m^2} &=&
\frac{e^{-\bar{\sigma}^2(t_1+t_2)}}{4\sinh^2|eB|(t_1+t_2)} \exp\left(-\frac{2y\sinh|eB|t_1
\sinh|eB|t_2}{\sinh|eB|(t_1+t_2)}\right)\nonumber\\
&\times&\left(1-\frac{2y\sinh|eB|t_1\sinh|eB|t_2}{\sinh|eB|(t_1+t_2)}\right).
\ea
Thus, we get the following representation for the correlators:
\ba
\Pi^{\sigma,\pi}(\Omega,k) &=& \frac{2\bar{\sigma}}{\pi l^2}\left(\tanh\frac{\bar\sigma}{2T}-1\right)
e^{-y}\left[\frac{1}{4\bar{\sigma}^2-\Omega^2} \frac{1\mp1}{2}-\int\limits_0^{\infty}dt_1dt_2
e^{(t_1+t_2)\Omega^2+2\bar{\sigma}\Omega(t_1-t_2)}\right.\nonumber\\
&\times&\left.\left(2|eB|ye^{-2|eB|(t_1+t_2)}e^{ye^{-2|eB|(t_1+t_2)}}
\frac{1\pm1}{2}+\Omega^2(e^{ye^{-2|eB|(t_1+t_2)}}-1)\frac{1\mp1}{2}\right)\right]\nonumber\\
&+&\frac{1}{\pi\sqrt{\pi} l^2}\int\limits_0^{\infty}\frac{dt_1dt_2}{\sqrt{t_1+t_2}}
e^{\frac{\Omega^2 t_1 t_2}{t_1+t_2} }\left(\frac{t_1+t_2+2t_1t_2\Omega^2\mp 2
\bar{\sigma}^2(t_1+t_2)^2}{2(t_1+t_2)^2}S(t_1,t_2)+2|eB|R(t_1,t_2)\right),
\label{correlators-intrepresentation}
\ea
which is convenient for expansions in $k^2$ and $\Omega^2$. It is also very useful for obtaining
the zero field limit, for that we get
\ba
\Pi^{\sigma,\pi}(\Omega,k)=\frac{1}{2\pi^{3/2}}\int\limits_{1/\Lambda^{2}}^{\infty}
\frac{d\rho\,e^{-\bar\sigma^{2}\rho}}{\rho^{3/2}}\int\limits_{0}^{1}dx\,e^{-(\mathbf{k}^{2}-\Omega^{2})
\rho x(1-x)}\left[\frac{3}{2}+\rho\left((\Omega^{2}-\mathbf{k}^{2})
x(1-x)\mp\bar\sigma^{2}\right)\right],
\ea
where an ultraviolet cutoff $\Lambda$ is introduced at the lower limit of integral.

It is obvious that the contribution of the first term in square brackets in
Eq.(\ref{correlators-intrepresentation}) is given by
\be
\Pi_{1}^{\sigma}(\Omega,k)=0,\,\,\,\,\,
\Pi_{1}^{\pi}(\Omega,k) = \frac{2}{\pi l^2}\frac{\bar{\sigma}\left(\tanh\frac{\bar\sigma}{2T}-1\right)}
{4\bar{\sigma}^2-\Omega^2}e^{-y}.
\ee
The contribution of other terms can be expanded in $y$ and $\Omega^{2}/|eB|$, and keeping only
the first order terms we get
\ba
\Pi^{\sigma}(\Omega,k)&=& \Pi_{1}^{\sigma}(\Omega,k)+ \frac{y }{\pi}\bar{\sigma}\left(1-
\tanh\frac{\bar\sigma}{2T}\right) +\frac{1}{\pi^{3/2}l}\Pi^{-}(\Omega,k),\\
\Pi^{\pi}(\Omega,k)&=& \Pi_{1}^{\pi}(\Omega,k) + \frac{1}{\pi^{3/2}l}\Pi^{+}(\Omega,k),
\ea
where
\be
\Pi^{\pm} = V^{\pm} + \frac{\Omega^2l^{2}}{4}Q^{\pm} + y P^{\pm},
\ee
and
\ba
V^{\pm} &=& \int\limits_{\epsilon}^{\infty}\frac{d\rho}{4\sqrt{\rho}}e^{-m^2\rho}
\left[(1\pm 2m^2\rho)\coth\rho+\frac{2\rho}{\sinh^2\rho}\right],\,\,\,
m^2=\bar{\sigma}^2l^{2},\\
Q^{\pm} &=& \frac{1}{2}\int\limits_{0}^{\infty}d\rho\,\,\sqrt{\rho} e^{-m^2\rho}
\left[\left(1\pm\frac{2m^2\rho}{3}\right)\coth\rho+\frac{2\rho}{3\sinh^2\rho}\right],\\
P^{\pm} &=&  \int\limits_{0}^{\infty}\frac{d\rho}{\sqrt{\rho}}e^{-m^2\rho}
\left[\frac{1\pm 2m^2\rho}{4\rho} \coth\rho+\frac{1}{\sinh^2\rho}\right](1-\rho\coth\rho).
\ea
The integral in the expression for $V^{\pm}$ is divergent and we regularized it by introducing
a lower limit cutoff $\epsilon=1/\Lambda^{2}l^{2}$. Finally, we get
\ba
V^{\pm} &=& \frac{3}{2\sqrt{\epsilon}} + \sqrt{\frac{\pi}{2}} \left[
\zeta\left(\frac{1}{2},1+\frac{m^2}{2}\right)- m^2 \zeta\left(\frac{3}{2},1+\frac{m^2}{2}\right)\frac{1\mp 1}{4}
+\frac{1\pm1}{2\sqrt{2}m}\right],\label{Vpm}\\
Q^{\pm} &=& \frac{1}{2}\sqrt{\frac{\pi}{2}} \left[
\zeta\left(\frac{3}{2},1+\frac{m^2}{2}\right)- m^2 \zeta\left(\frac{5}{2},1+\frac{m^2}{2}\right)\frac{1\mp 1}{4}
+\frac{1\pm1}{m^{3}\sqrt{2}}\right],\label{Qpm}\\
P^{+}&=&-\sqrt{\frac{\pi}{2}}\frac{1}{\sqrt{2}m},\label{Pplus}\\
P^{-} &=& -\sqrt{\frac{\pi}{2}}\left\{m^{2}\zeta\left(\frac{1}{2},1+\frac{m^{2}}{2}\right)
+\frac{m^{4}}{2}\zeta\left(\frac{3}{2},1+\frac{m^{2}}{2}\right)+\sqrt{2}m\right\},
\label{Pminus}
\ea
where $\zeta(s,v)$ is the generalized zeta function.


\begin{thebibliography}{99}

\bibitem{Jackiw} R. Jackiw, Phys. Rev. D {\bf 29}, 2375 (1984);
I. Affleck, Nucl. Phys. B {\bf 265}, 409 (1986);
A. Kovner and B. Rosenstein, Phys. Rev. B {\bf 42}, 4748 (1990);
G.W. Semenoff and L.C.R. Wijewardhana, Phys. Rev. B {\bf 45}, 1342 (1992);
R. MacKenzie, P.K. Panigrahi, and R. Sakhi, Phys. Rev. B {\bf 48}, 3892 (1993).

\bibitem{Semenoff-graphene}G.W.~Semenoff, Phys. Rev. Lett. {\bf53}, 2449 (1984).

\bibitem{d-wave}M.~Franz and Z.~Te$\breve{s}$anovi$\acute{c}$, Phys. Rev. Lett. {\bf87}, 257003 (2001);
I.~Herbut, Phys. Rev. B {\bf66}, 094504 (2002).

\bibitem{top-insulators}C.~L.~Kane and E.~J.~Mele, Phys. Rev. Lett. {\bf95}, 146802 (2005);
M.~Z.~Hasan and C.~L.~Kane, Rev. Mod. Phys. {\bf82}, 3045 (2010).

\bibitem{opt-lattice}S.-L.~Zhu, B.~Wang, and L.-M.~Duan, Phys. Rev. Lett. {\bf98}, 260402 (2007);
A.~Singha, M.~Gibertini, B.~Karmakar, S.~Yuan, M.~Polini, G.~Vignale, M.I.~Katsnelson, A.~Pinczuk,
L.~N.~Pfeiffer, K.~W.~West, and V.~Pellegrini, Science {\bf332}, 1176 (2011).

\bibitem{Novoselov} K.S. Novoselov, A.K. Geim, S.V. Morozov, D. Jiang, Y. Zhang,
S.V. Dubonos, I.V. Grigorieva, and A.A. Firsov, Science {\bf 306}, 666 (2004).

\bibitem{graphene-review}V.P.~Gusynin, S.G.~Sharapov, and J.P.~Carbotte, Int.
J. Mod. Phys. B{\bf21}, No.27, 4611 (2007);
A.H. Castro-Neto, F. Guinea, N.M.R. Peres, K.S. Novoselov,
and A.K. Geim, Rev. Mod. Phys. {\bf 81}, 109 (2009);
D.S.L. Abergel, V. Apalkov, J. Berashevich, K. Ziegler, and T. Chakraborty,
Advances in Physics {\bf 59}, 261 (2010);
N.~M.~R.~Peres, Rev. Mod. Phys. {\bf82}, 2673 (2010);
V.N.~Kotov, B.~Uchoa, V.M.~Pereira, A.~H.~Castro Neto, and F.~Guinea, arXiv:1012.3484v1
[cond-mat.str-el] (to appear in Rev. Mod. Phys.);
S.~Das Sarma, S.~Adam, E.H. Hwang, and E.~Rossi, Rev. Mod. Phys. {\bf83}, 407 (2011).

\bibitem{Alicea} J.~Alicea and M.P.A.~Fisher, Phys. Rev. B {\bf 74}, 075422 (2006).

\bibitem{NJL} Y. Nambu and G. Jona-Lasinio, Phys. Rev. {\bf 122}, 345 (1961).

\bibitem{Kleinert} H. Kleinert, {\sl On the Hadronization of Quark Theories},
Lectures presented at the Erice Summer Institute 1976; in: {\sl Understanding
the Fundamental Constituents of Matter}, A. Zichichi (ed.), Plenum Press,
New York, 1978, p. 289.

\bibitem{Volkov} M.K. Volkov, Ann. Phys. (N.Y.) {\bf 157}, 282 (1984).

\bibitem{Hatsuda} T. Hatsuda and T. Kunihiro, Phys. Lett. B {\bf 145}, 7 (1984).

\bibitem{Shankar} R. Shankar, Rev. Mod. Phys. {\bf 66}, 129 (1994); J. Polchinski,
in {\sl Proceedings of the 1992 TASI}, edited by J. Harvey and J. Polchinski
(World Scientific, Singapore, 1993), hep-th/9210046.

\bibitem{catalysis1}
V.P.~Gusynin, V.A.~Miransky, and I.A.~Shovkovy,
Phys. Rev. Lett. {\bf 73}, 3499 (1994);
Phys.  Rev.  D {\bf 52}, 4718 (1995).

\bibitem{catalysis2} V.P.~Gusynin, V.A.~Miransky, and I.A.~Shovkovy, Phys. Lett.
B {\bf 349}, 477 (1995).
\bibitem{catalysis3}I.V.~Krive and S.A.~Naftulin, Phys. Rev. D {\bf46}, 2737 (1992);
K.G. Klimenko, Z. Phys. C {\bf54}, 323 (1992); Theor. Math. Phys. {\bf89}, 1161 (1992);
C.N. Leung, Y.J. Ng, and A.W. Ackley, Phys. Rev. D {\bf54}, 4181 (1996);
K. Farakos and N.E. Mavromatos, Int. J. Mod. Phys. B {\bf12}, 809 (1998);
G. Jona-Lasinio and F. M. Marchetti, Phys. Lett. B {\bf459},  208 (1999);
G. Jona-Lasinio, Progr. Theor. Phys. {\bf124}, 731 (2010);
E.J. Ferrer and V. de la Incera, Phys. Lett. B {\bf481}, 287 (2000).

\bibitem{UFG}V.P.~Gusynin, Ukr. J. Phys. {\bf45}, 603 (2000).

\bibitem{Semenoff} G.~W.~Semenoff, I.~A.~Shovkovy, and L.~C.~R.~Wijewardhana, Phys. Rev.
D {\bf60}, 105024 (1999).

\bibitem{catalysis-QED} V.P.~Gusynin, V.A.~Miransky, and I.A.~Shovkovy, Phys. Rev.
D {\bf 52}, 4747 (1995);
Phys. Rev. Lett. {\bf83}, 1291 (1999);
Nucl. Phys. B {\bf 563}, 361 (1999);
V.P.~Gusynin and A.V.~Smilga, Phys. Lett. B {\bf450}, 267 (1999);
C.N. Leung and S.-Y. Wang, Nucl. Phys. B {\bf747}, 266 (2006);
E. Rojas, A. Ayala, A. Bashir, and A. Raya, Phys. Rev. D {\bf77}, 093004 (2008).

\bibitem{QCD}I.A.~Shushpanov and A.V.~Smilga, Phys. Lett. B {\bf402}, 351 (1997);
V.A.~Miransky and I.A.~Shovkovy, Phys. Rev. D {\bf66}, 045006 (2002);
R.~Gatto and M.~Ruggieri, Phys. Rev. D {\bf83}, 034016 (2011);
A.J.~Mizher, E.S.~Fraga, and M.N.~Chernodub, arXiv:1103.0954 [hep-ph].

\bibitem{holography}V.G.~Filev, C.V.~Johnson, and J.P.~Shock, JHEP {\bf08}, 013 (2009);
V.G.~Filev and R.C.~Raskov, Adv. High Energy Phys. {\bf2010}, 473206 (2010);
V.G.~Filev and D.~Zoakos, JHEP {\bf08}, 022 (2011);
J.L.~Davis, H.~Omid, and G.W.~Semenoff, JHEP {\bf09}, 124 (2011).

\bibitem{Schrieffer-book} J.R. Schrieffer, {\sl Theory of superconductivity} (W.A. Benjamin, 1964).

\bibitem{Cooper} L.N. Cooper, Phys. Rev. {\bf 104}, 1189 (1956).

\bibitem{Rivista} P.I. Fomin, V.P. Gusynin, V.A. Miransky, and Yu.A. Sitenko,
Riv. Nuovo Cimento {\bf 6}, No. 5, 1 (1983).

\bibitem{Zeldovich} Ya.B. Zeldovich and V.N. Popov, Sov. Phys. Usp. {\bf 14}, 673 (1972).

\bibitem{Greiner} W. Greiner, B. Muller, and J. Rafelski, {\sl Quantum Electrodynamics
of Strong Fields} (Springer-Verlag, Berlin, 1985).

\bibitem{3G} O.V. Gamayun, E.V. Gorbar, and V.P. Gusynin, Phys. Rev. B {\bf 83}, 235104 (2011);
Ukr. J. Phys. {\bf56}, 688 (2011).

\bibitem{Incera}E.J.~Ferrer, V.P.~Gusynin, and V.~de la Incera, Eur. Phys. J. B {\bf33}, 397 (2003).

\bibitem{Schwinger} J.S. Schwinger, Phys. Rev. {\bf 82}, 664 (1951).

\bibitem{Schwinger1} J.S. Schwinger, J. Math. Phys. {\bf 2}, 407 (1961).

\bibitem{Keldysh} L.V. Keldysh, Sov. Phys. JETP {\bf 20}, 1018 (1965).

\bibitem{real-time-review}A.J.~Niemi and G.W.~Semenoff,  Ann. Phys. {\bf152}, 105 (1984);
K.C. Chou, Z.B. Su, B.L. Hao, and L. Yu, Phys. Rep. {\bf 118}, 1 (1985);
N.P. Landsman and Ch.G. van Weert, Phys. Rep. {\bf 145}, 141 (1987).

\bibitem{Fu} W. Fu, D. Huang, and F. Wang, Nucl. Phys. A {\bf849}, 203 (2011).

\bibitem{Pyatkoskiy}P.K.~Pyatkovskiy and V.P.~Gusynin, Phys. Rev. B {\bf83}, 075422 (2011).


\end{thebibliography}
\end{document}